**Effects of edge chemistry doping on graphene nanoribbon mobility**

Yijian Ouyang[1], Stefano Sanvito[2] and Jing Guo[1, *]

[1]Department of Electrical and Computer Engineering, University of Florida, Gainesville, FL 32611, USA

[2] Physics Department, Trinity College, Dublin 2, Ireland

**Abstract**

Doping of semiconductor is necessary for various device applications. Exploiting chemistry at its reactive edges was shown to be an effective way to dope an atomically thin graphene nanoribbon (GNR) for realizing new devices in recent experiments. The carrier mobility limited by edge doping is studied as a function of the GNR width, doping density, and carrier density by using *ab initio* density functional and parameterized tight binding simulations combined with the non-equilibrium Green's function formalism for quantum transport. The results indicate that for GNRs wider than about 4nm, the mobility scales approximately linearly with the GNR width, inversely proportional to the edge doping concentration and decreases for an increasing carrier density. For narrower GNRs, dependence of the mobility on the GNR width and carrier density can be qualitatively different.

* Corresponding author: email guoj@ufl.edu, phone 1-352-3920940, fax 1-352-3928381

# 1. Introduction

The high mobility of graphene, which is a monolayer of carbon atoms packed into a two-dimensional (2D) honeycomb lattice, has stimulated strong interest in high-performance graphene device applications [1-3]. Graphene nanoribbons (GNR) have been obtained by patterning graphene into narrow strips. The 2D graphene does not have a bandgap. A bandgap, however, can be opened by quantum confinement in the width direction of a GNR [4-7]. The chemically active edges can be engineered for various functionalities despite that experiment control over edge chemistry remains challenging at current stage. A recent work has shown that chemical reactions can turn a p-type GNR into an n-type GNR as it is annealed in ammonia [8]. The impact of edge dopants on the carrier mobility, however, is not yet clear. In this study the edge-dopant-limited mobility is investigated, which has not been covered by previous studies on the GNR mobility [9], [10].

In this work, chemistry of different dopants in a GNR is captured by using *ab initio* density functional theory (DFT) simulations. For n-type doping, substitution of edge carbon atoms by nitrogen atoms is considered, and passivation of edge carbon atoms by oxygen atoms is considered for p-type doping [8]. The quantum transport equation is solved in the non-equilibrium Green's function (NEGF) formalism with Hamiltonians in tight binding (TB) bases parameterized from *ab initio* Hamiltonian and overlap matrices [11]. The dependences of the edge-dopant-limited mobility on the dopant species, doping concentration, GNR channel width, and carrier density are investigated. The study shows that the quasi-one-dimensional channel and resonant scattering play an important role in determining the edge-dopant-limited GNR mobility.

## 2. Approach

### 2.1. *Ab initio* DFT simulation

The electronic transmission as a function of energy is simulated for a GNR with a single N or O edge dopant as respectively shown in Fig. 1(a) or (b) by SMEAGOL program [12], which combines DFT method and NEGF method. The simulated doped channel has 5 GNR unit cells with one dopant at the edge of the middle unit cell and the contacts are semi-infinite GNR leads. Structure relaxation is first performed for both the channel and the contacts by the *ab initio* density functional package SIESTA with a criterion of 0.04eV/Å [13]. Due to the short range of the dopant perturbation, the Hamiltonians of the two unit cells at the two ends of the channel supercell are found to approximately recover to that of a perfect GNR unit cell. A double zeta polarized (DZP) basis set is used. The energy cut-off is 200 Ryd. Exchange and correlation are treated in local density approximation. Based on the relaxed channel and contact structures, electronic transmission is calculated by SMEAGOL. The Hamiltonian and overlap matrices of the channel and leads in the DZP basis are obtained, which are used for TB parameterization as described next.

### 2.2. TB parameterization

Extraction of mobility requires simulations of hundreds of randomly generated doped GNR structures at each channel length as described below. Hence *ab initio* simulations are computationally too expensive. A TB parameterization is therefore attractive for significantly improving computational efficiency. An orthogonal $p_z$ orbital TB description has been extensively used in simulating undoped CNT and GNR electronic structures at low energy [14], [15]. For GNRs, a correction to Hamiltonian elements due to the edge bond relaxation is necessary to obtain a good agreement with the band structures calculated by *ab initio* simulations [15]. After including a dopant, the band structure is neither a useful nor a meaningful criterion to test whether a TB parameterization is good or not due to the breaking of translational symmetry. Instead the transmission through a doped channel connected to two semi-infinite GNR leads can be used to judge the validity of the parameterization [16]. It has been shown that only adding changes to the diagonal elements of the Hamiltonian in the orthogonal $p_Z$ TB basis is sufficient to reproduce the transmission calculated by *ab initio* transport simulations in a previous study of CNTs with a B or N substitutional dopant [16] . As shown in Fig. 2(a) and (b), we found that this method works well for the nitrogen edge doping whose structure is shown in Fig. 1(a).

For oxygen edge doping in Fig. 1(b), the orthogonal $p_Z$ TB basis set with only on-site changes, however, is found not able to produce transmissions that agree with those calculated by SMEAGOL. We instead found that by truncating the *ab initio* Hamiltonian and overlap matrices of the DZP basis into the first zeta $p_Z$ orbital of the DZP basis set only, the transmissions calculated in the reduced basis set agree with the SMEAGOL results as shown in Fig. 2(c) and (d). The Hamiltonian and overlap matrices in the first zeta $p_Z$ orbital basis are essentially a subset of the original *ab initio* basis set. Thus this Pz basis is non-orthogonal and has coupling between atoms beyond first nearest neighbors. We keep the non-orthogonality and limit the interaction up to third nearest neighbors. Perturbations to both on-site elements and hopping elements are included. Despite the complexity in the TB parameterization for oxygen doping, the size of the non-orthogonal basis, which includes one orbital per atom (hydrogen atoms are not in the Hamiltonian anymore), is the same as the orthogonal $p_Z$ TB basis for nitrogen edge doping.

With computational efficiency improved significantly by TB parameterization, the resistance of a randomly generated GNR structure with multiple dopants as shown in Fig. 1(c) can be calculated in the TB basis set. Cautions, however, must be taken when the configurations of multiple dopants are randomly generated. The TB bases are derived from *ab initio* simulations of a single dopant. When there are multiple dopants, it is assumed that each dopant causes a perturbation of Hamiltonian and overlap matrices in a way similar to that by a single dopant. Therefore the dopants should not be too close to each other and a superposition of perturbations can hold. The interaction between two closely spaced dopants increases the total energy of a GNR, which is not favored by chemical reactions [17]. In this study any two of the dopants are not allowed to sit at the same edge if they are in one unit cell or in two neighboring unit cells.

## 2.3. Mobility extraction

After constructing the channel Hamiltonian, the retarded Green's function of the channel at energy $E$ is calculated as,

$$G(E) = [(E + i0^+)I - H - \Sigma_1 - \Sigma_2]^{-1} \quad (1)$$

where $H$ is the Hamiltonian matrix in the orthogonal (non-orthogonal) $p_Z$ TB basis for nitrogen (oxygen) doped channels, and $\Sigma_1$ ($\Sigma_2$) is the self-energy due to the semi-infinite dopant-free source (drain) lead. For oxygen doped channels, the identity matrix $I$ is replaced by the overlap matrix. The electronic transmission per spin through the channel is calculated as

$$Tr(E) = \text{Tr}[\Gamma_1(E)G(E)\Gamma_2(E)G^+(E)] \tag{2}$$

where $\Gamma_{1,2}(E) = i(\Sigma_{1,2} - \Sigma_{1,2}^+)$ is the broadening function of the source/drain lead and $G^+$ is the advanced Green's function. NEGF formalism can rigorously treat the transport in the presence of multiple dopant potentials. Other non-idealities for transport such as phonon scattering and structural defects are neglected since this study focuses on the dopant-limited mobility. With the transmission, the resistance can be computed as,

$$R = 1/G,$$

$$G = \frac{2q^2}{h} \int dE\, Tr(E) \left( -\frac{\partial f(E, E_F, T)}{\partial E} \right), \tag{3}$$

where $G$ is the conductance and $f$ is the Fermi distribution which is a function of energy $E$, Fermi energy $E_F$ and temperature $T$. The factor of 2 counts the spin degeneracy. An $E_F$ can be related to a specified 2D carrier density, which is the carrier per GNR length divided by the GNR width. For a given channel length, width, edge doping concentration, and 2D carrier density, 300 trial simulations with randomly generated dopant configurations are performed to obtain an averaged resistance, $<R>$, which can be partitioned into a contact resistance $R_0$ and a channel resistance $R_{ch}$,

$$<R> = R_0 + R_{ch} \approx R_0 + (1/\sigma)L_{ch}, \tag{4}$$

where the conductivity $\sigma$ is computed from the slope of a linear fitting to the $<R>$ vs. $L_{ch}$ curve. The electron mobility is calculated as

$$\mu_n = \frac{\sigma}{qn_{2D}W} \tag{5}$$

where $n_{2D}$ is the 2D electron density and $W$ is the channel width. The hole mobility is obtained by changing $n_{2D}$ to $p_{2D}$, the 2D hole density. The extraction of channel resistivity should be performed in the linear regime of the <*R*> vs. $L_{ch}$ curve to avoid the onset of localization.

## 3. Simulation Results

### 3.1. Electronic transmissions

First we examine the *ab initio* transport simulation results of GNRs with a single dopant as shown in Fig. 1(a) and (b). The TB transport simulations are also performed for the same structures to check the validity of the parameterization. The simulated transmissions as a function of energy for nitrogen edge doping are displayed in Fig. 2(a) and (b), for small bandgap armchair GNRs (AGNR) in the *n*=3*q*+2 group and semiconducting AGNRs in the *n*=3*q* group, respectively. The results of the *n*=3*q*+1 group are not shown because they are qualitatively similar to those of the *n*=3*q*+2 group. It is noticed that there is a qualitative difference between the transmissions of the valence subbands and conduction subbands. In both Fig. 2(a) and (b), oscillations (peaks and dips) appear in the transmissions of the conduction subbands whereas the transmissions of the valence subbands monotonically increase as |*E*| increases. As the channel width becomes smaller, the transmission peaks and dips occur at larger energies due to the larger energy spacing between subbands.

Figure 2(c) and (d) show the results of oxygen edge doping. Oscillations of transmissions similar to nitrogen edge doping are observed as well except that the conduction and valence subbands are switched. The energies at which the oscillation dips are observed can be explained by resonant backscattering, and they align with the peak values of the local density-of-states (not shown here), manifesting the quasi-local states due to the quantum confinement by the dopant potential. For n-type doping by nitrogen, the ionized dopant (positively charged) induces a potential well for electrons and resonant back scattering takes place in the conduction subbands. In contrast, for p-type doping by oxygen, the ionized dopant (negatively charged) induces a potential well for holes, and resonant backscattering takes place in the valence subbands.

Agreement between the *ab initio* results and the TB results in Fig. 2 indicates the validity of the TB parameterization. The TB approach is attractive for enabling efficient simulations of longer GNRs with multiple dopants as shown in Fig. 1(c) to extract dopant-limited mobility values. The parameterized change of the diagonal entry of the Hamiltonian at the nitrogen substitution site is about -4 eV for n-type doping, and that at the carbon atom site which is passivated by an oxygen atom is about 1eV for p-type doping. The smaller magnitude of the perturbation in p-type doping is because an oxygen atom passivates a carbon atom while a nitrogen atom substitutes for a carbon atom. As a result, in subsequent simulations, the hole mobility of an oxygen doped AGNR is consistently larger than the electron mobility of a nitrogen doped AGNR with the same channel width, doping concentration and carrier density.

**3.2. Channel width dependence**

Figure 3 indicates the simulated GNR resistance as a function of the channel length, in which each data point (a circle) is obtained by averaging over 300 randomly generated doped channel configurations. The resistance varies linearly with the channel length, which indicates diffusive nature of the transport in the simulated channel length regime and localization does not occur. The dopant-limited mobility is extracted from the slope of the linear fitting to the simulated data points, whose extrapolation at $L_{ch}$=0 is confirmed to agree with the expected ballistic resistance for the simulated carrier density which determines $E_F$.

The dependence of the dopant-limited mobility on the AGNR width is examined next. In Fig. 4(a) and (b), the mobility vs. channel width curves are plotted for nitrogen doping and oxygen doping, respectively. For each type of dopants, two curves are plotted, one for the *n=3q* group and the other for the *n=3q+2* group. Regardless of the GNR width, the 2D carrier density is $1 \times 10^{13}$ m$^{-2}$ and the edge doping concentration is 0.02. The Fermi energy $E_F$ determines the 2D charge density, which is the carrier per unit length divided by the GNR width, and the edge doping concentration is defined as the ratio of the number of dopants to the number of edge carbon atoms. For the simulated GNRs wider than about 4nm, the dopant-limited mobility increases approximately linearly as the GNR width increases for a fixed edge doping concentration. It indicates a smaller edge effect on the transport properties of a wider GNR. As the channel width increases, the probability for a carrier being close to the edges lowers and the matrix element of the dopant potential between the initial and final wave states decreases, carrier transport is less perturbed and the mobility increases.

For a GNR narrower than about 4 nm, non-monotonic behaviors are observed in Fig. 4, which can be explained by the position of the Fermi energy with regard to the transmission dips and peaks caused by resonant scattering. Take the curve for the *n=3q+2* group of nitrogen doping in Fig. 4(a) as an example. For a fixed 2D carrier density of $1 \times 10^{13}$ cm$^{-2}$, we found that the Fermi energy is about 0.3eV above (below) the charge neutral point for electron (hole) conduction regardless of the channel width. The Fermi energy $E_F \approx 0.3$eV is near the first transmission peak of the small channel width (e.g. *n*=8 as shown in Fig. 2(a) and the top diagram of the inset in Fig.4(a)). As the channel width increases, the position of first transmission dip is shifted toward the Fermi energy. The transmission at $E_F$ decreases and reaches the minimum value when $E_F$ is right aligned to the dip as shown in the middle diagram of the inset in Fig.4(a), which results in a decrease of the conductance as indicated by Eq. (4), and thereby the mobility. As the channel width further increases, the first dip moves away from the Fermi energy and the mobility increases indicated by the bottom diagram of the inset in Fig.4(a) (Figure 2 shows the transmission curve of a single dopant, but that of multiple dopants is qualitatively similar to that of a single dopant in terms of the energies of resonant transmission dips due to backscattering[18]). Resonant backscattering dips also exist in the transmission of higher subbands, but the dips in higher subbands are much shallower and narrower than that in the lowest subband which reduces transmission close to zero. In the presence of room temperature thermal broadening, the transmission dips in higher subbands are not strong enough to cause non-monotonic dependence of the mobility on the GNR width.

### 3.3. Doping concentration dependence

Next we investigate the dependence of the dopant-limited mobility on the edge doping concentration at a fixed 2D carrier density of $1 \times 10^{13} \text{cm}^{-2}$. As the doping concentration increases, the density of dopants increases and the carriers get more frequently backscattered so the mobility decreases as shown by the solid lines in Fig 5(a) and (b) for both nitrogen and oxygen doping respectively. The dashed lines are fitting lines proportional to $\sim 1/N_{D,A}$ where $N_D$ ($N_A$) is the nitrogen (oxygen) edge dopant concentration. The inversely proportional dependence of the mobility on the doping concentration is qualitatively the same as in silicon [19]. In Fig. 5(b) modeled for oxygen doping, the curve of $n$=15 AGNR is very close to that of $n$=30 AGNR in log scales, due to the non-monotonic dependence on the width for narrow GNRs as shown for the $n=3q$ group in Fig. 4(b).

**3.4. Carrier density dependence**

The Fermi energy and carrier density can be modulated by gating in a GNR device. We next study the dependence of mobility on the carrier density. Figure 6 plots the dopant-limited hole mobility as a function of the 2D carrier density with an oxygen edge doping concentration of 0.02 for three different GNR widths. The simulated $n$=15, 30, and 66 AGNRs have widths of 1.9nm, 3.7nm, and 8.1nm respectively, and the simulated 2D carrier density varies from $1 \times 10^{12}$ to $2 \times 10^{13}$cm$^{-2}$. Qualitatively different trends are observed for AGNRs with different widths.

Figure 6 shows that for a narrow GNR with $n$=15, the mobility slightly increases as the carrier density increases. In the simulated carrier density range, the conductance increases monotonically as the carrier density increases as shown in Fig. 7(a). The transmission dip in the lowest subband occurs below $E_F$ =-0.44eV that results in a carrier density of $2 \times 10^{13}$cm$^{-2}$ and therefore does not have an effect in the simulated range of the carrier density. Furthermore, only the lowest subband is relevant in the simulated hole density range. The band-structure-limited carrier velocity increases and the density-of-states decreases at the Fermi energy as the hole density increases. The increase of the conductivity slightly outpaces the increase of the hole density in the simulated hole density range, which results in a slight increase of the mobility as determined by Eq. (5).

In contrast, Figure 6 indicates that the mobility decreases as the hole density increases for a wider $n$=66 AGNR with a width of 8.1nm. To understand this phenomenon, the transmission as a function of energy is plotted in Fig. 7(b) with the Fermi energies at three simulated densities denoted. Several differences from the $n$=15 AGNR case are noted. First, as the hole density increases, the Fermi energy moves into a range with several resonant backscattering dips that lower the transmission and conductance. Second, four subbands are involved in the simulated hole density range. The band-structure-limited carrier velocity is zero and the van Hove singularity of density-of-states exists at the bottom of each subband, which results in a rapid increase of the hole density as $E_F$ moves near the bottom of any subband. As a result, the increase of the carrier density is more dominant than the increase or even decrease of the conductivity as $E_F$ moves away from $E$=0. Therefore the mobility decreases as the hole density increases. The non-monotonic dependence for the $n$=30 AGNR with a width of 3.7nm as shown in Fig. 6 can be explained by a combined effect. For low hole density, only the first subband is involved in transport and the transmission dip is out of the relevant energy range, the mobility increases as the hole density increases. Further increase of the hole density results in a decrease of the mobility, due to the higher subband transport and transmission dips caused by resonant backscattering.

4. **Conclusions and Discussions**

Edge chemistry provides a unique way to dope graphene nanostructures. While other scattering mechanisms could also degrade intrinsically high mobility of graphene, scattering by edges has been considered as the dominant mechanism for limiting carrier mobility in a narrow graphene nanoribbon. The edge-dopant-limited mobility of both n-type and p-type GNRs with edge doping is examined in this work. The nitrogen edge doping results in oscillations in the conduction subband transmissions whereas the oxygen edge doping results in oscillations in the valence subband transmissions due to resonant scattering. Resonant scattering plays an important role in the dependence of the dopant-limited mobility on the channel width, edge doping concentration, and 2D carrier density. The mobility approximately scales linearly with the GNR width when the width *W*>4nm and varies non-monotonically with the width for narrower GNRs. The inversely proportional dependence of the mobility on the doping concentration is qualitatively similar to that in silicon. In the carrier density range of interest, the mobility slightly increases as the 2D carrier density increases for an AGNR with *W*~2nm whereas the mobility decreases as the carrier density increases for an AGNR with *W*~8nm. The effect of edge doping on GNR mobility can be incorporated with those due to other scattering mechanisms (edge roughness, adsorbate impurity, and phonon) to first order through the Matthiessen's rule.

**Acknowledgements**

It is our pleasure to thank Prof. Hongjie Dai and Dr. Xinran Wang of Stanford University for helpful discussions. This work is supported by ONR, NSF, and ARL.

**References**

[1] K. S. Novoselov et al., "Electric Field Effect in Atomically Thin Carbon Films," *Science*, vol. 306, no. 5696, pp. 666-669, Oct. 2004.

[2] Y. Zhang, Y. Tan, H. L. Stormer, and P. Kim, "Experimental observation of the quantum Hall effect and Berry's phase in graphene," *Nature*, vol. 438, no. 7065, pp. 201-204, Nov. 2005.

[3] C. Berger et al., "Electronic Confinement and Coherence in Patterned Epitaxial Graphene," *Science*, vol. 312, no. 5777, pp. 1191-1196, May. 2006.

[4] M. Y. Han, B. Ozyilmaz, Y. Zhang, and P. Kim, "Energy Band-Gap Engineering of Graphene Nanoribbons," *Physical Review Letters*, vol. 98, no. 20, p. 206805, May. 2007.

[5] Z. Chen, Y. Lin, M. J. Rooks, and P. Avouris, "Graphene nano-ribbon electronics," *Physica E: Low-dimensional Systems and Nanostructures*, vol. 40, no. 2, pp. 228-232, Dec. 2007.

[6] X. Li, X. Wang, L. Zhang, S. Lee, and H. Dai, "Chemically Derived, Ultrasmooth Graphene Nanoribbon Semiconductors," *Science*, vol. 319, no. 5867, pp. 1229-1232, Feb. 2008.

[7] X. Wang, Y. Ouyang, X. Li, H. Wang, J. Guo, and H. Dai, "Room-Temperature All-Semiconducting Sub-10-nm Graphene Nanoribbon Field-Effect Transistors," *Physical Review Letters*, vol. 100, no. 20, pp. 206803-4, May. 2008.

[8] X. Wang et al., "N-Doping of Graphene Through Electrothermal Reactions with Ammonia," *Science*, vol. 324, no. 5928, pp. 768-771, May. 2009.

[9] A. Betti, G. Fiori, G. Iannaccone, and Y. Mao, "Physical insights on graphene nanoribbon mobility through atomistic simulations," in *Electron Devices Meeting (IEDM), 2009 IEEE International*, pp. 1-4, 2009.

[10] T. Fang, A. Konar, H. Xing, and D. Jena, "Mobility in semiconducting graphene nanoribbons: Phonon, impurity, and edge roughness scattering," *Physical Review B*, vol. 78, no. 20, p. 205403, Nov. 2008.

[11] S. Datta, *Quantum Transport: Atom to Transistor*. Cambridge, UK: Cambridge University Press, 2005.

[12] S. Sanvito, C. J. Lambert, J. H. Jefferson, and A. M. Bratkovsky, "General Green's-function formalism for transport calculations with spd Hamiltonians and giant magnetoresistance in Co- and Ni-based magnetic multilayers," *Physical Review B*, vol. 59, no. 18, p. 11936, May. 1999.

[13] J. M. Soler et al., "The SIESTA method for ab initio order-N materials simulation," *Journal of Physics: Condensed Matter*, vol. 14, no. 11, p. 2745.

[14] R. Saito, M. S. Dresselhaus, and G. Dresselhaus, *Physical properties of carbon nanotubes*. Imperial College Press, 1998.

[15] Y. Son, M. L. Cohen, and S. G. Louie, "Energy Gaps in Graphene Nanoribbons," *Physical Review Letters*, vol. 97, no. 21, pp. 216803-4, Nov. 2006.

[16] C. Adessi, S. Roche, and X. Blase, "Reduced backscattering in potassium-doped nanotubes: Ab initio and semiempirical simulations," *Physical Review B (Condensed Matter and Materials Physics)*, vol. 73, no. 12, pp. 125414-5, Mar. 2006.

[17] T. B. Martins, R. H. Miwa, A. J. R. da Silva, and A. Fazzio, "Electronic and

       Transport Properties of Boron-Doped Graphene Nanoribbons," *Physical Review Letters*, vol. 98, no. 19, p. 196803, May. 2007.
[18] T. Markussen, R. Rurali, A. Jauho, and M. Brandbyge, "Scaling Theory Put into Practice: First-Principles Modeling of Transport in Doped Silicon Nanowires," *Physical Review Letters*, vol. 99, no. 7, p. 076803, 2007.
[19] E. Conwell and V. F. Weisskopf, "Theory of Impurity Scattering in Semiconductors," *Physical Review*, vol. 77, no. 3, p. 388, Feb. 1950.

**Figure captions**

**Figure 1.** Atomic structures of *n*=15 AGNR super-cells with a single dopant for *ab initio* simulations and a channel segment with multiple dopants for TB transport simulations. (a) N-type edge doping by substituting a nitrogen atom for an edge carbon atom. (b) P-type edge doping by passivating an edge carbon atom with an oxygen atom. (c) A channel segment with multiple N dopants.

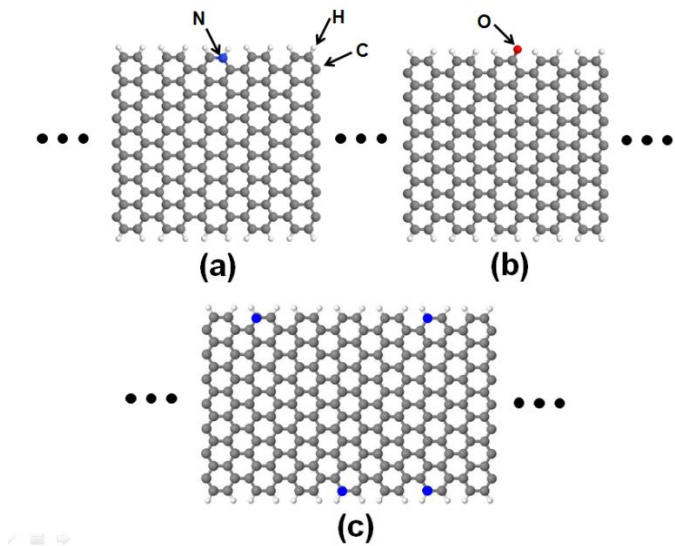

**Figure 2.** Transmission as a function of energy computed with DZP basis used in SMEAGOL simulations (solid lines) and with the parameterized TB bases (dashed lines) in the presence of a single N dopant in (a) $n$=8, 14 AGNRs and (b) $n$=9, 15 AGNRs or a single O dopant in (c) $n$=8, 14, 20 AGNRs and (d) $n$=9, 15 AGNRs. The middle of the bandgap is defined as $E$=0.

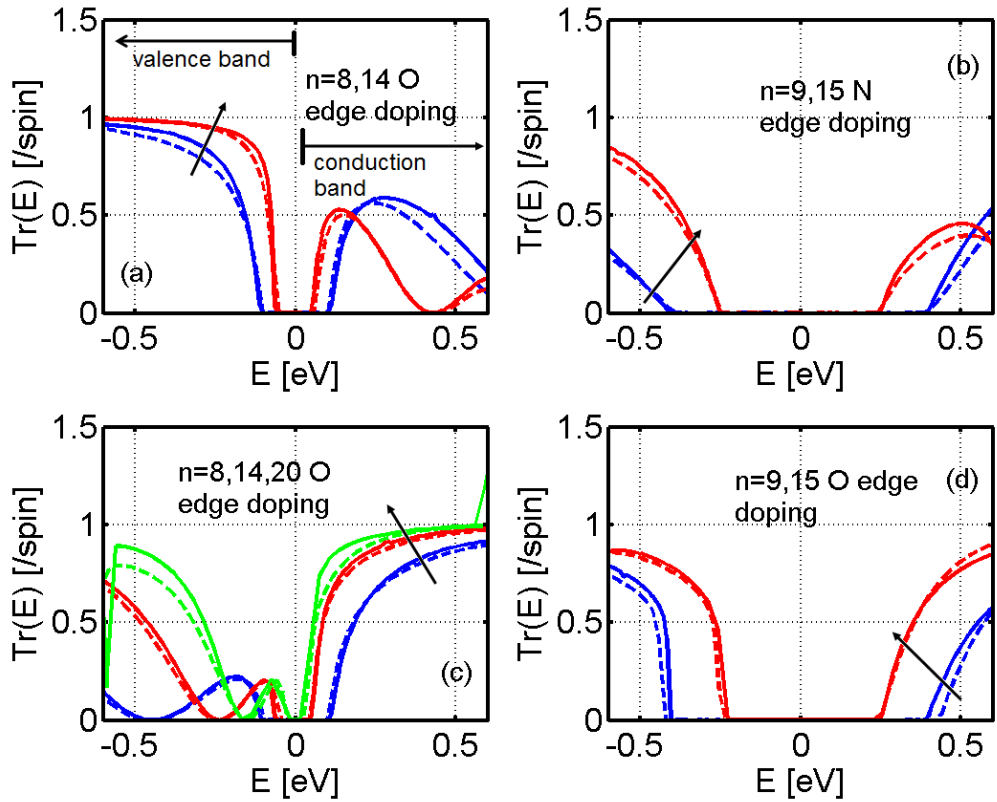

**Figure 3.** Resistance as a function of the channel length for $n=66$ N doped AGNRs with an edge doping concentration of 0.02 (the ratio of the number of dopants to the number of edge carbon atoms). The dashed curve is the linear fitting of the simulated results marked by the circles. The square shows the value of the ballistic resistance of 3 subbands ($h/6e^2$), which agrees well with the extrapolated resistance at $L_{ch}=0$ from the dashed line. The Fermi energy $E_F \approx 0.33$eV, which results in an equivalent 2D electron density of $1 \times 10^{13}$cm$^{-2}$.

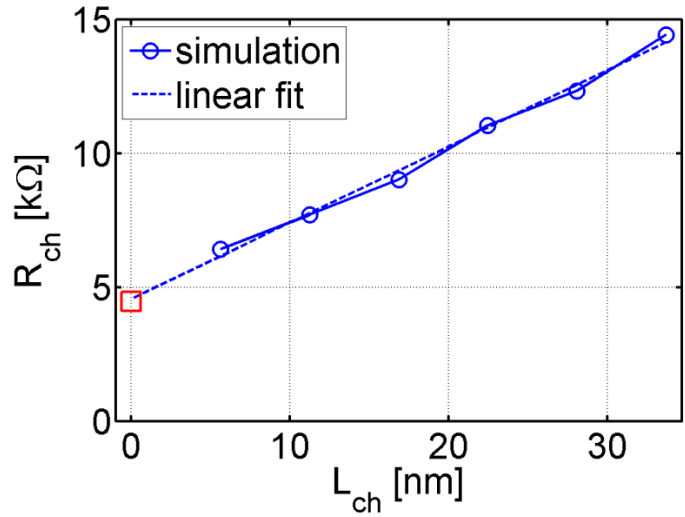

**Figure 4.** Edge-dopant-limited mobility as a function of the channel width for (a) nitrogen edge doping and (b) oxygen edge doping at a 2D carrier density of $1 \times 10^{13} cm^{-2}$ and an edge doping concentration of 0.02. The inset of Fig.4(a) schematically shows the position of the Fermi level with regard to the transmission spectra at three channel widths $w_A$, $w_B$ and $w_C$ to explain the non-monotonic behavior of the mobility as a function of the channel width for n=3q+2 group.

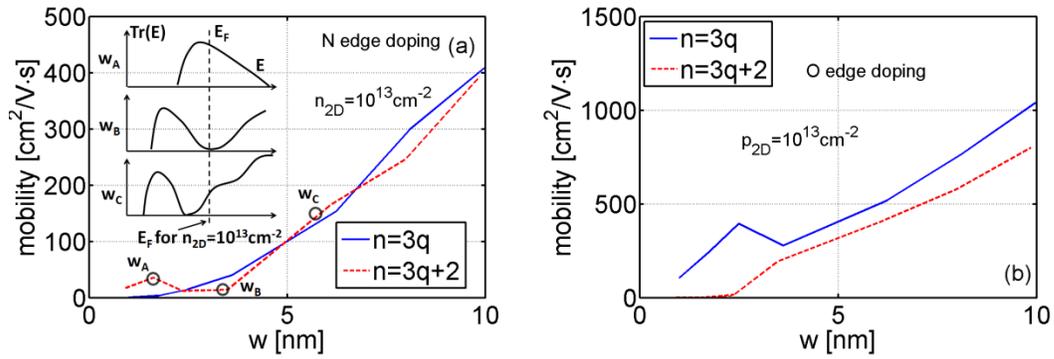

**Figure 5.** Edge-dopant-limited mobility as a function of the edge doping concentration for (a) nitrogen edge doping and (b) oxygen edge doping with a 2D carrier density of $1 \times 10^{13} \text{cm}^{-2}$.

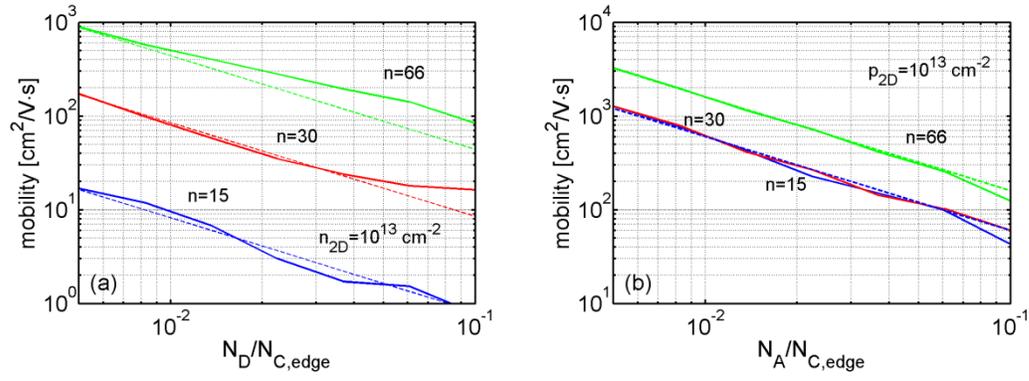

**Figure 6.** Edge-doping-limited mobility as a function of the 2D carrier density with oxygen edge doping at three channel widths. The edge doping concentration is 0.02. The triangular and square symbols are marked at the carrier densities of $1\times10^{12}$, $1\times10^{13}$, and $2\times10^{13}$cm$^{-2}$ which correspond to the Fermi energies labeled respectively as $E_{F1}$, $E_{F2}$, and $E_{F3}$ in Fig.7.

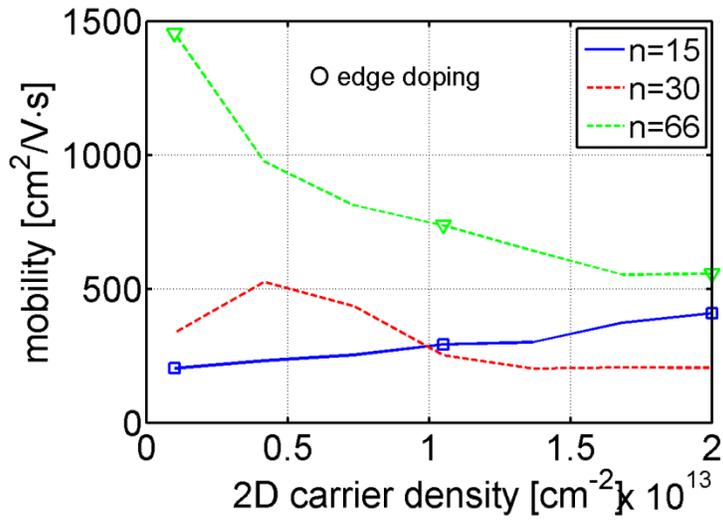

**Figure 7.** Transmission as a function of energy of (a) an n=15 AGNR and (b) an $n=66$ AGNR for oxygen edge doping. The red dashed curves are the step-wise perfect transmissions. The vertical lines labeled as $E_{F1}$, $E_{F2}$, and $E_{F3}$ indicate the Fermi energies for the carrier density $1\times10^{12}$, $1\times10^{13}$, and $2\times10^{13}\text{cm}^{-2}$ respectively.

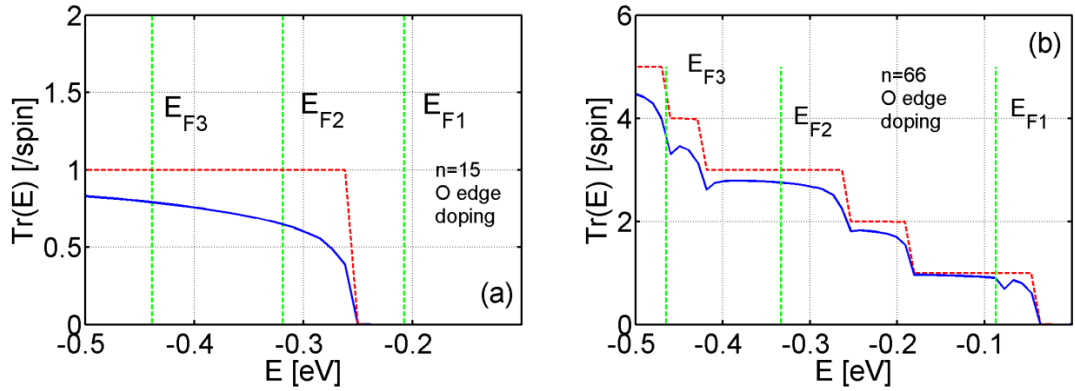